\documentclass[a4paper,12pt]{article} 
\usepackage{epsfig}
\usepackage{graphicx}

\newcommand{\beq}{\begin{eqnarray}}
\newcommand{\eeq}{\end{eqnarray}}

\begin{document}
\pagestyle{plain}

%
%
\title{
{\Large \bf 
Beamstrahlung monitoring of the beam beam effects at the Linear Collider
}
}
 
\author{
{\bf 
Nicolas Delerue\footnote{\tt nicolas@post.kek.jp}
 and
 Toshiaki Tauchi\footnote{\tt toshiaki.tauchi@kek.jp}
} \\ 
\em High Energy Accelerator Research Organization (KEK),\\
\em 1-1 Oho, Tsukuba Science City, 305-0801 Ibaraki-ken, Japan
}

\date{
August 2004
}

\maketitle

%
\begin{center}
{\bf 
At the Linear Collider mismatches between the two beams will result in
an intense beamstrahlung. We have studied how this beamstrahlung would
evolve as a function of the offset between the two beams and we suggest
 ways of monitoring it.
} \\ 
\end{center}


%
\section{Beamstrahlung at the Linear Collider.}
%

At the Linear Collider the colliding beams will have a vertical size
of less than a few nanometers. With such a small size any misalignement of
the final focus magnets will lead to an offset between the two beams
at the interaction point (IP). As the magnetic field in each bunch
will be of the order of kilo-Teslas, the misalignement of the two beams
will make that each beam will travel accross the strong magnetic dipole
created by the other and thus produce an intense synchrotron radiation
called, in this case, beamstrahlung~\cite{Yokoya:1991qz}.

We have used CAIN~\cite{Chen:1995jt} to evaluate the intensity of the
beamstrahlung as a function of the offset between the two beams. The
beam parameters used are those of the JLC as
described~\cite{unknown:2003na} but our conclusions are valid for any
linear collider operationg around 500~GeV and would be similar at an
energy of 1~TeV.
In
this study we have neglected the coherent beamstrahlung (CB) as the
energy of the CB photons is well below the energies discussed below.

%
%
\begin{figure}[p]
\centering\epsfig{file=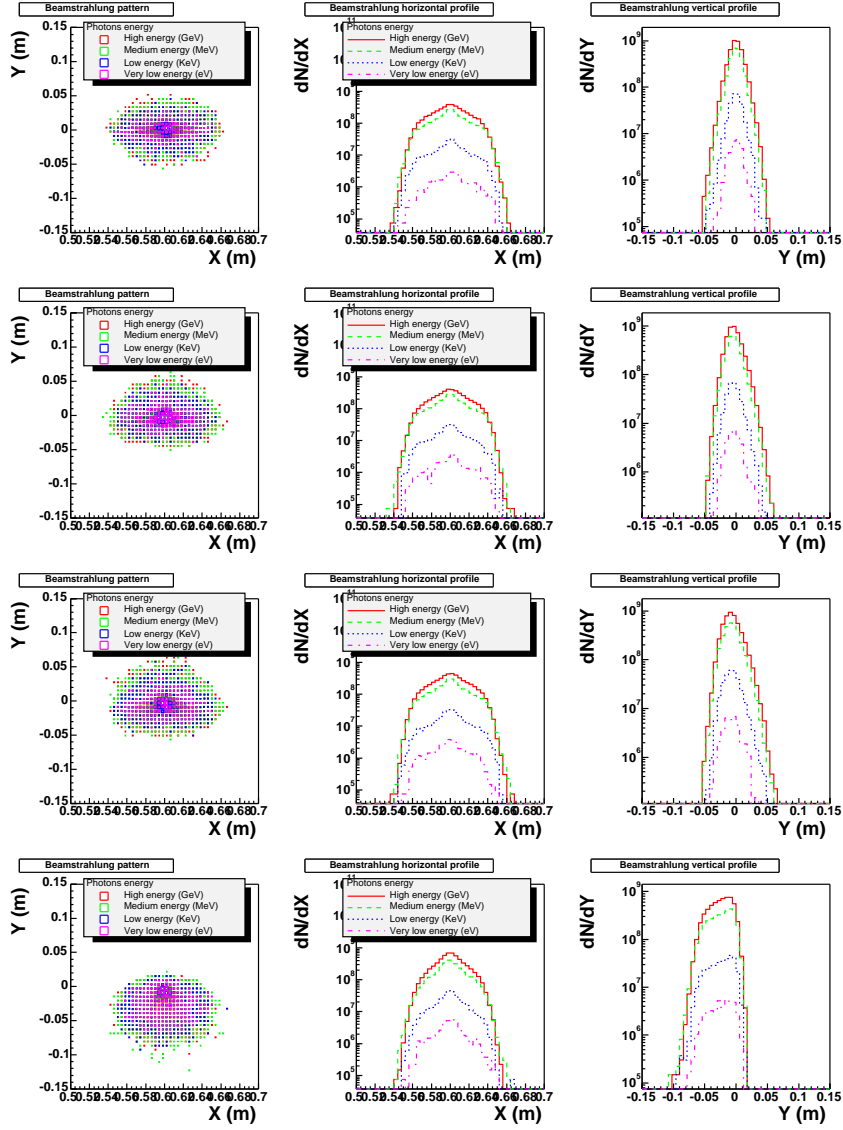,width=12.cm}
\caption{Beamstrahlung spot 200 meters
away from the IP for different beam offset. For 4 different beam beam
vertical offsets: (from top to bottom) 0, 0.5, 1 and 10 times the size of the
beams. For each offset the leftmost figure is the beamstrahlung
pattern, the figure in the middle gives the horizontal profile of
the pattern and the rightmost figure gives the vertical profile of the
pattern. The four different colors correspond to the four energy range
(eV, KeV, MeV and GeV). When the vertical offset changes, the pattern,
as well as its vertical profile change but the horizontal profile
remains constant.}
\label{fig:beamspot} 
\end{figure}

The figure~\ref{fig:beamspot} shows the beamstrahlung spot 200 meters
away from the IP for different vertical offset between the two beams. As one can see there is a
clear dependance of the beamstrahlung pattern with the beam offset. 

\begin{figure}[p]
\centering\epsfig{file=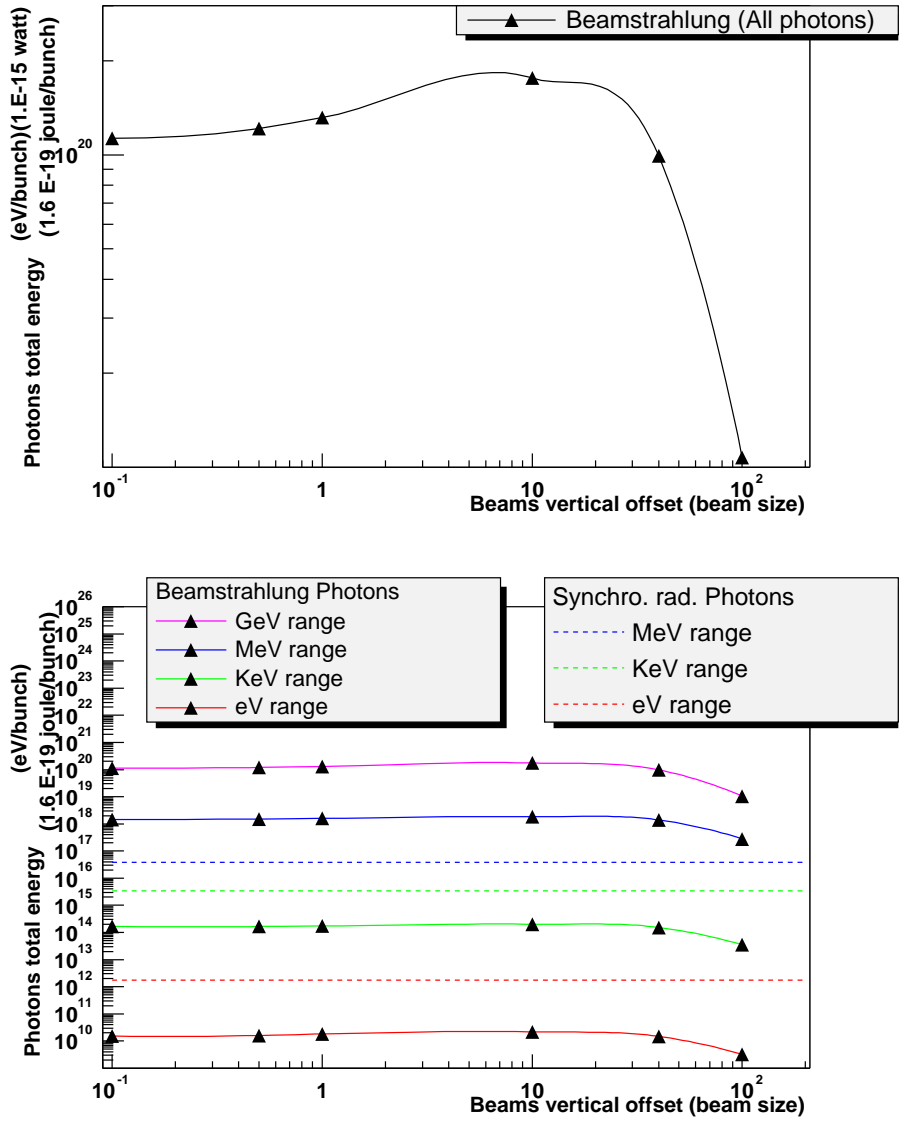,width=12.cm}
\caption{Power  delivered  by the beamstrahlung as a function of the
  vertical beam beam offset. The upper plot shows the total power
  delivered and the lower plot shows the power delivered for each
  energy range. On the lower plot the power delivered by the
  synchrotron radiation produced by the final focusing is also
  shown.}
\label{fig:power} 
\end{figure}

\begin{figure}[p]
\centering\epsfig{file=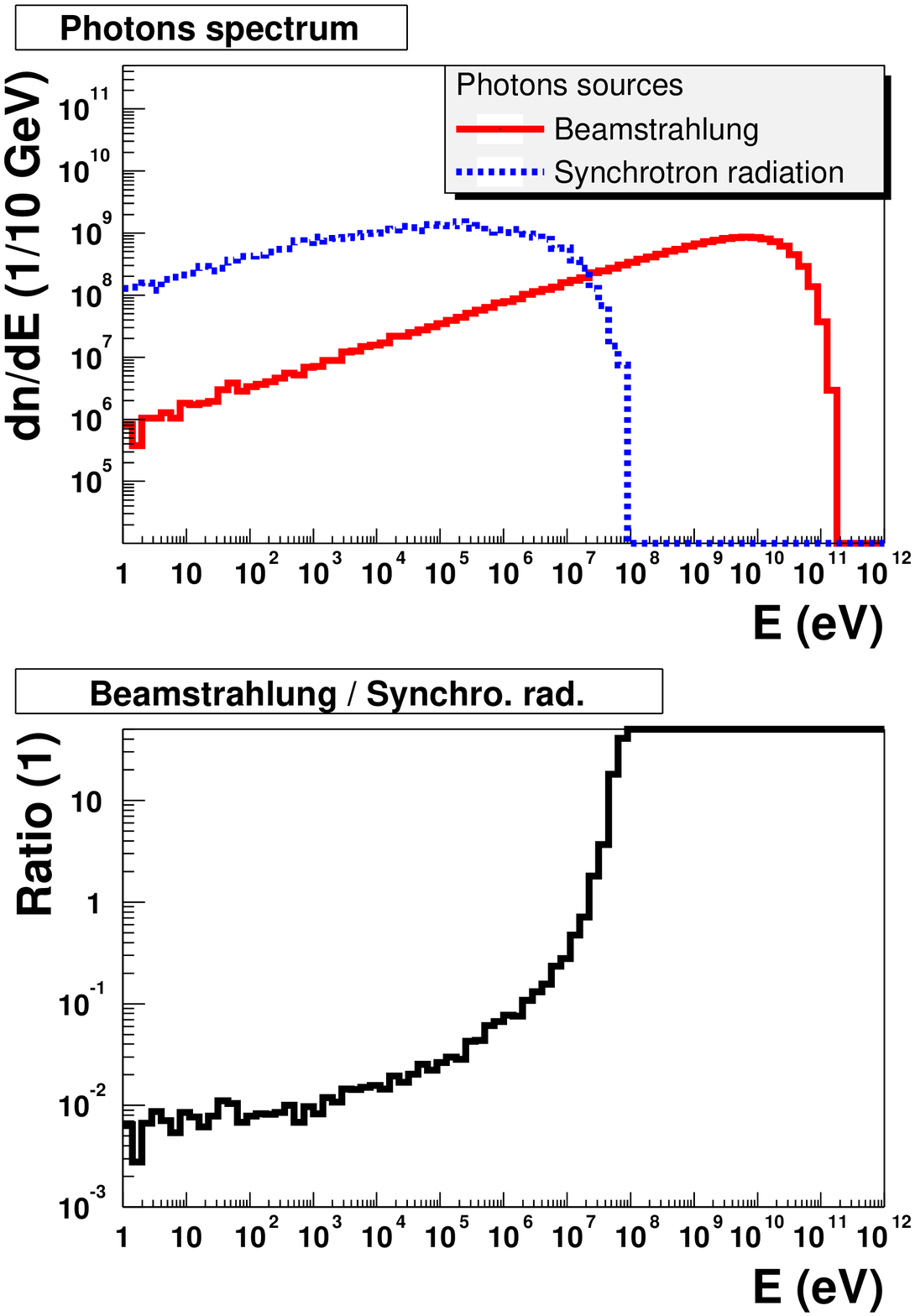,width=14.cm}
\caption{Energy spectrum of the beamstrahlung photons and of the
  synchrotron radiation photons produced by the final focusing (upper
  plot) and ratio between these two sources (lower plot).}
\label{fig:ratio} 
\end{figure}

The figure~\ref{fig:power} shows the total power per bunch of the
beamstrahlung. As one can see the power also changes with the beams
offset. It is also important to note on
this figure that at high energy (MeV and above) the beamstrahlung intensity is higher than
the synchrotron radiation generated by the bending of the electrons
during the final focusing whereas at lower energy (KeV and below) the
beamstrahlung photons are drowned in a synchrotron radiation
background. A more detailed comparison of the beamstrahlung and
synchrotron radiation spectrums is shown on figure~\ref{fig:ratio},
upper plot.

%
%
%

%
%

As the beam offset will have an important influence on the luminosity of
the Linear Collider, monitoring the beamstrahlung pattern would
provide usefull information on the beam offset.

%
\section{Possible beamstrahlung monitors}
%

One of the main challenge faced by a breamstrahlung monitor will be
the intensity of the beamstrahlung. As one can see on
figure~\ref{fig:power}, the total energy of the radiations emitted by
each bunch will be of the order of a dizains of joules per bunch. As
there will be more than 10~000 bunch per second the total power
delivered by the beamstrahlung will be a few kilowatts. A
conventionnal imaging system would not be able to stand such power
and would be instantly destroyed if it was placed in the beam line.
Thus alternative solutions have to be proposed to monitor the beamstrahlung.

\subsection{Using the low energy photons reflected by a mirror}

The beamstrahlung radiation will be absorbed by a water tank.
As the walls of the tank will be made of metal, one can imagine to
polish this metal to make it reflective. The reflectivity of most
metals is low for very high energy photons whereas it becomes higher
for photons of a few eV~\cite{Caso:1998txp152}. Thus the most energetic part of the spectrum
would be absorbed by the water tank whereas the lower part of the
spectrum would be partly reflected toward an imaging device, allowing
the observation of the beamstrahlung pattern in the optical spectrum. One of the challenge
faced by this imaging device would be to extract the variation of the 
beamstrahlung pattern from the huge synchrotron radiation
background (2 orders of magnitude bigger, see figure~\ref{fig:ratio},
lower plot). Thus the imaging device would have to be sensitive to
intensity variation of the order of a few percents.
To increase the spatial resolution of the imaging device one could
give a spherical shape to the mirror rather than a flat one, thus increasing the
size of the beamstrahlung pattern on the imaging device.

%
%
\begin{figure}[htbp]
\begin{tabular}{p{7cm}p{7cm}}
\centering\epsfig{file=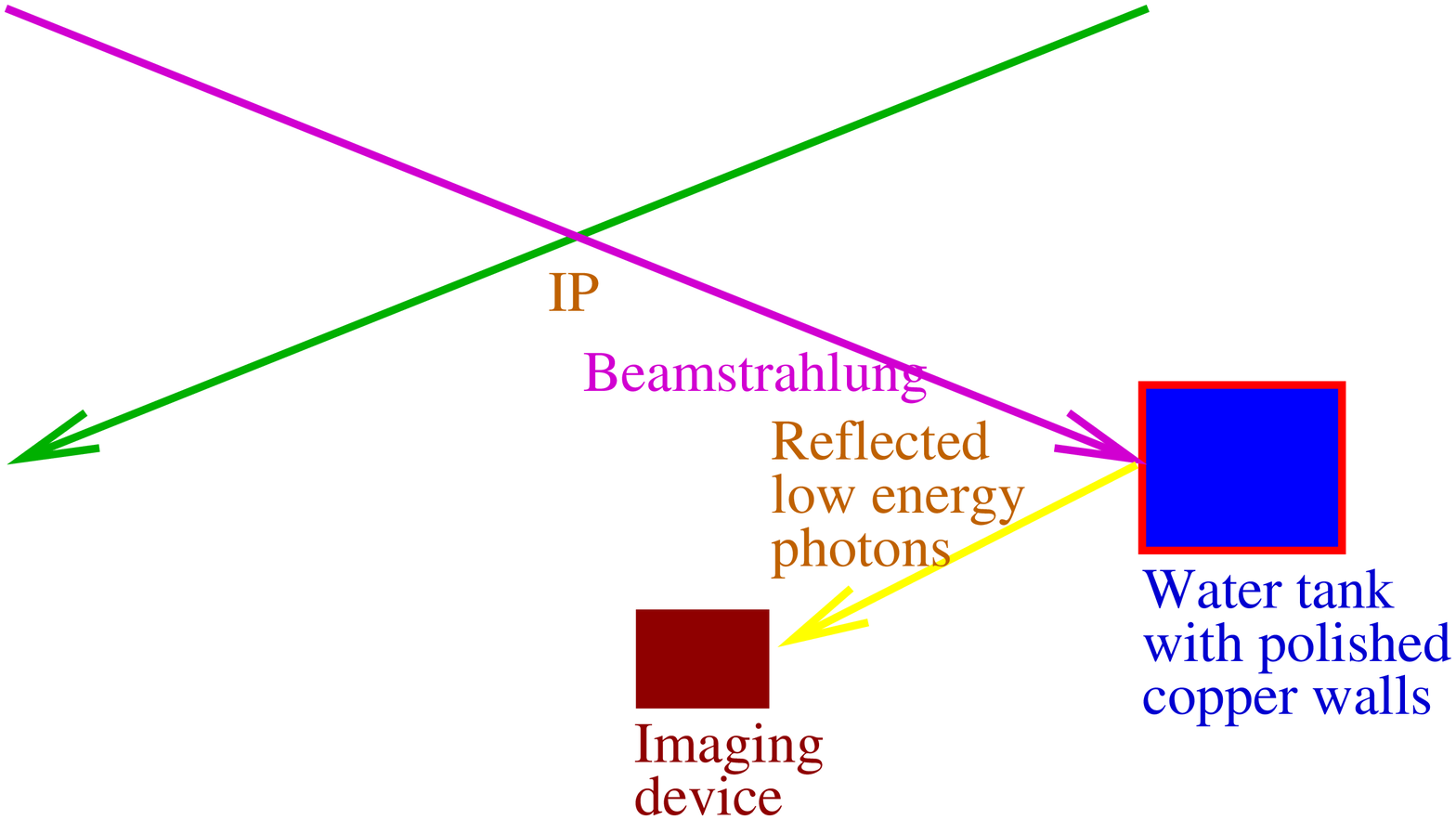,width=7.cm}
\caption{The walls of the water tank used to absord the radiations
  could be used as a mirror to reflect the lower part of the radiation
  spectrum.}
\label{fig:Mirror} 
&
\centering\epsfig{file=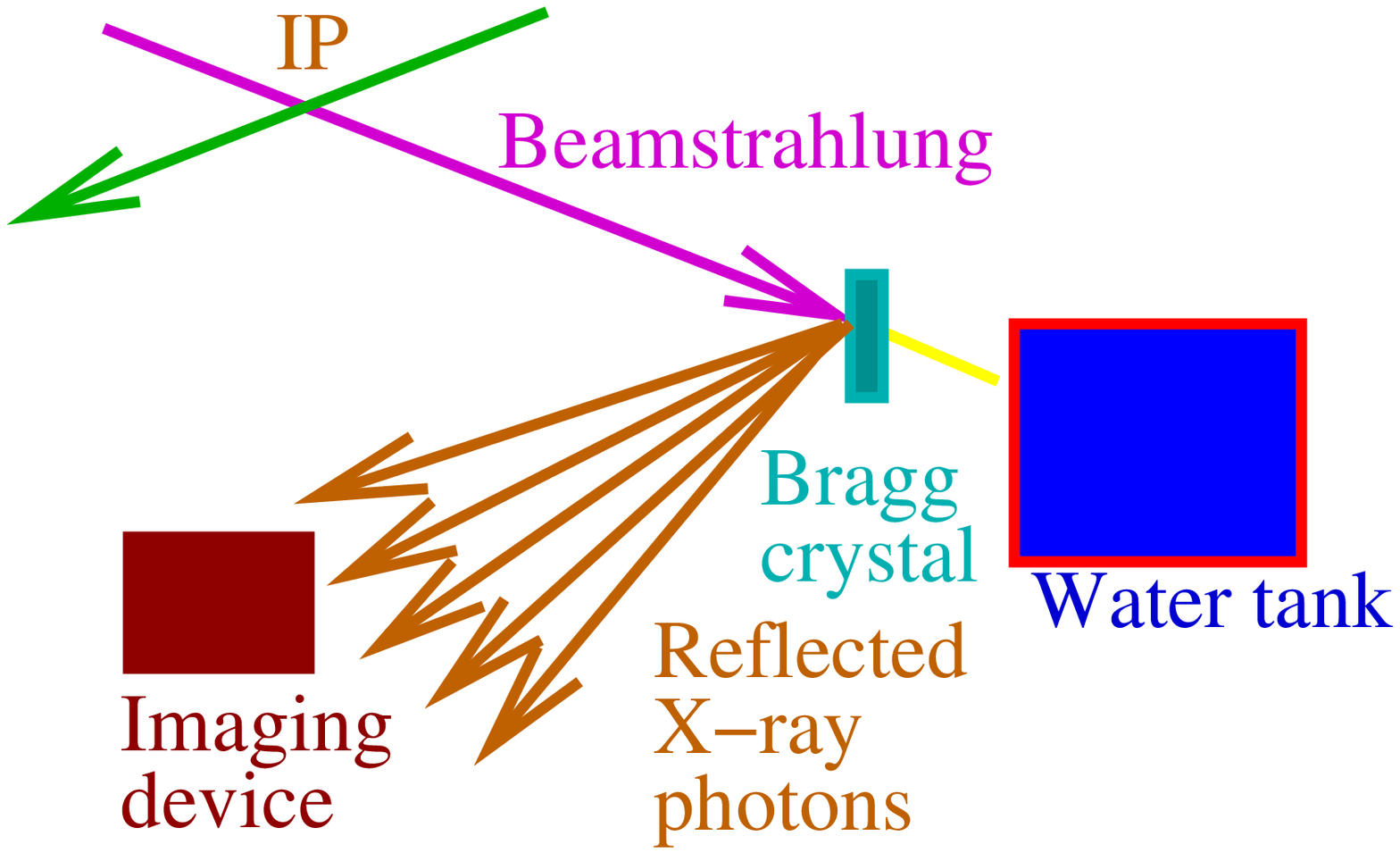,width=6.cm}
\caption{A Bragg crystal could be used to select only hard X-rays
  where the signal to noise ratio is better.}
\label{fig:Bragg} 
\end{tabular}
\end{figure}

\subsection{Using a Bragg crystal to select thard X-rays}

To increase the signal to noise ratio, one may want to select hard X-rays
photons (in the hundreds of KeV range) instead of optical photons. In the KeV
range the signal to noise ratio is slightly better than in the optical
range (as shown on figure~\ref{fig:ratio}, lower plot), thus detecting
changes in the beamstrahlung pattern would be easier.

To select photons in the KeV range a Bragg crystal could be used. This
Bragg crystal would reflect photons of different wavelength in
different directions, thus allowing to easily select only one small
energy range as shown on figure~\ref{fig:Bragg}.

The main difficulty of this layout is that the Bragg crystal would
have to be inserted directly in the beamstrahlung flux and thus would
be exposed to a very intense radiation (a few kilowatts for a 1mm
thick crystal).

\subsection{Ionization Chamber}

As seen in figure~\ref{fig:power}  (top), the total photon energy depends on the beam offset,
especially in case of large offset where the deflection angle would be too
small for the fast feedback.  An ionization chamber can be a candidate
detector for measurement of the photon flux.  The chamber has a 1mm thick
gap filled with gas such as Helium, whose pressure can be less than 1 atm.
Front plate of the chamber must have enough thickness for absorption of
synchrotron radiations, e.g. a few cm thick copper plate.  The area covers
the photon distribution shown in figure~\ref{fig:beamspot}, e.g. 20cm diameter at 200m from IP.
The photons convert into electron-positron pairs in the front plate.  The
electrons and positrons ionize the gas. The ionized electrons are detected
at the back plate with readout electronics. The gap thickness shall be
optimized with drift time.  Since typical drift velocity is 4cm/$\mu$sec
with $\sim$2kV/cm electric field in the gap, the drift time is estimated to
be 40nsec.  Since number of ionized electrons can be $10^9$/bunch in this
configuration, no additional amplification would be necessary.  Segmented
back plate must be very desired option for position measurement.  The
segmentation size of 1cm$\times$1cm seems to be enough as seen in figure~\ref{fig:beamspot}.

\subsection{Electron Wire}

%
%
\begin{figure}[htbp]
\centering\epsfig{file=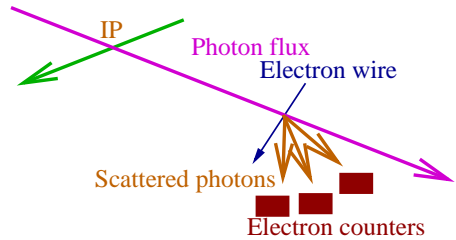,width=6.cm}
\caption{An electron wire could scan the radiation beam. Electrons
  would be scattered in different directions depending on the energy
  of the photons the have hit. Electron counters would be used to
  measure the scattered electrons' energy and reconstruct the beam spectrum.}
\label{fig:electronWire} 
\end{figure}

To measure the high energy part of the spectrum without inserting any
device in the beamline, one could try to use an
``electron wire'' to scan the beam: a thin beam of medium energy (a few
MeV) electrons would be sent accross the beamstrahlung beam. The
electrons of the beam would be scattered by the photons. The angle
with which the electrons would be scattered would be proportionnal to
the photons' energy. By installing electrons counters at different
angles one would be able to measure the photon beam spectrum at
different energies. 
The most forward counters (observing electrons that have been
scattered by highly energetic photons) would observe almost only the
beamstrahlung spectrum.
The observed spectrum would evolve when the beams offset
changes.

%
\section{Conclusion}
%

The beamstrahlung radiation generated during the beam crossing at the
IP will carry some information on the beam beam offset. Monitoring the
beamstrahlung pattern would thus allow to extract these
information. We have proposed different setup that would allow the
monitoring of the the beamstrahlung pattern despite its high power
and the high background.

%
\section{Acknowledgements}
%

We would like to thank Kaoru Yokoya for the many useful discussions we
had on this topic. We would also like to thank Tomomi Ohgaki who
provided us the synchrotron radiation simulations.

One of the authors (ND) would like to thank JSPS 
for funding his stay in Japan under contract P02794.

\bibliographystyle{myunsrt}
\bibliography{biblio}

\end{document}